\documentclass[nofootinbib,prd,twocolumn,showpacs,showkeys,preprintnumbers]{revtex4-1}
\usepackage{hyperref,amssymb,amsmath,mathrsfs,bm,graphicx}

\usepackage{enumitem}
\usepackage{color}
\begin{document}

\title {A Methodological Framework for Solving Einstein’s Equations in Axially Symmetric Spacetimes}
\author{ J. Ospino and J.L. Hern\'andez-Pastora}
\email{j.ospino@usal.es}
\affiliation{Departamento de Matem\'atica Aplicada and Instituto Universitario de F\'isica
Fundamental y Matem\'aticas, Universidad de Salamanca, Salamanca 37007, Spain}

\author{A.V. Araujo-Salcedo}
\email{aaraujo@campestre.edu.co}
\affiliation{Centro de Estudios, Gimnasio Campestre, Bogot\'a, Colombia}

\author{L.A. N\'u\~nez}
\email{ lnunez@uis.edu.co}
\affiliation{ Escuela de F\'isica, Universidad Industrial de Santander, Bucaramanga, Colombia\\
 Departamento de F\'isica, Universidad de Los Andes, M\'erida, Venezuela}

\begin{abstract}
This work presents a novel methodology for deriving stationary and axially symmetric solutions to Einstein's field equations using the 1+3 tetrad formalism. This approach reformulates the Einstein equations into first-order scalar equations, enabling systematic resolution in vacuum scenarios. We derive two distinct solutions in polar and hyperbolic geometries by assuming the separability of a key metric function. Our method reproduces well-known solutions such as Schwarzschild and Kerr metrics and extends the case of rotating spacetimes to hyperbolic configurations. Additionally, we explore the role of Killing tensors in enabling separable metric components, simplifying analyses of geodesic motion and physical phenomena. This framework demonstrates robustness and adaptability for addressing the complexities of axially symmetric spacetimes, paving the way for further applications to Kerr-like solutions in General Relativity.
\end{abstract}
\maketitle

\section{Introduction}
The Event Horizon Telescope (EHT) has revolutionised our understanding of axially symmetric spacetimes by providing groundbreaking observational evidence. Its iconic images of the supermassive black holes in M87 and Sagittarius A* unveiled shadow-like features that align with theoretical predictions based on the Kerr metric \cite{AkiyamaEtal2019A, AkiyamaEtal2019B}. These observations represent the first direct confirmation of an event horizon's existence, offering crucial constraints on the spin and mass of black holes \cite{AkiyamaEtal2022A}. A key coming goal of the EHT is to measure the properties of emission rings, apparent shadows, and other observables that are directly influenced by the spacetime characteristics of M87* and Sgr A* \cite{AkiyamaEtal2024A}.

These groundbreaking images have sparked significant scientific discussions, particularly regarding the implications of two key astrophysical vacuum solutions to Einstein's field equations: the Schwarzschild and Kerr metrics. Notably, some observations from the object in $Sgr A*$ appear to conflict with the predictions of the Kerr solution \cite{AkiyamaEtal2022A}. Researchers have explored these discrepancies, with several studies attempting to explain the geodesic motion of the $S$-stars using the classical rotating Kerr solution \cite{PerlickEtal2022}. Alternatively, \cite{YounsiEtal2023} proposed two distinct spacetime geometries: one derived from a modified gravity theory and another that deviates parametrically from the Kerr solution, reducing to Kerr spacetime when the deviation parameters vanish.

Decades before this observational boom, a considerable effort has been dedicated to developing methods for deriving axisymmetric and stationary solutions to Einstein's vacuum field equations. Comprehensive reviews of these techniques and their results can be found in \cite{Quevedo1990, StephaniEtal2006} and the references therein. Early in the history of General Relativity, in  1917, Weyl \cite{Weyl1917} derived the general family of axisymmetric static solutions. It was just one year after Schwarzschild's pioneering spherically symmetric solution \cite{Schwarzschild1999} was published. Decades later, in 1985, Gutsunayev and Manko \cite{GutsunayevManko1985, DenisovaKhakimovManko1994} reformulated this family of solutions using a different representation of Weyl's work. However, it was Roy Kerr who, in 1963, discovered the first physically realistic stationary axisymmetric solution \cite{Kerr1963}, marking a pivotal moment, leading to the development of various techniques and encouraging significant efforts in the relativistic community to derive exact solutions to Einstein's vacuum equations.

Early solution-generating methods incorporate the Kerr-Schild ansatz \cite{Kerr1963}, complex transformations \cite{NewmanJanis1965}, and approaches based on Hamilton-Jacobi separability \cite{Carter1968}. Other notable contributions include techniques by Newman, Tamburino, and Unti \cite{NewmanTamburinoUnti1963}, Kinnersley \cite{Kinnersley1969}, and Plebanski and Demianski \cite{PlebanskiDemianski1976}.

Much of the progress in generating axially symmetric solutions is credited to Ernst's scheme \cite{Ernst1968A, Ernst1968B}, which reformulated the Einstein system and introduced two powerful generating techniques. These methods were later expanded and unified by Kinnersley \cite{Kinnersley1973, Kinnersley1977}. Building on Ernst's compact field equation framework, Tomimatsu and Sato \cite{TomimatsuSato1972, TomimatsuSato1973}, along with Yamazaki and collaborators \cite{YamazakiHori1977, Yamazaki1977A, Yamazaki1977B, YamazakiHori1978}, developed new solutions by specifying particular functional forms for the transformation of the Ernst potential.

Modern generating techniques often involve Lie group or B\"acklund transformations. For example, Cosgrove utilised these methods to derive generalised Tomimatsu-Sato solutions \cite{Cosgrove1977A, Cosgrove1977B}, while Maison \cite{Maison1979} studied an infinite-dimensional group and the Geroch group \cite{Geroch1971, Geroch1972}. Although the Geroch group does not yield asymptotically flat solutions, Kinnersley and Chitre \cite{KinnersleyChitre1977, KinnersleyChitre1978A, KinnersleyChitre1978B, KinnersleyChitre1978C} identified infinitesimal subgroups that preserve asymptotic flatness. Using the Zipoy-Vorhees metric \cite{Zipoy1966}, they generated solutions that include the NUT class and the Tomimatsu-Sato metric as particular cases. Later, Belinskii and Zakharov \cite{BelinskiiZakharov1978} formulated a linear eigenvalue problem equivalent to the nonlinear field equations, solvable via the inverse scattering method. This approach was further developed by researchers such as Sibgatullin \cite{Sibgatullin1984} and Harrison \cite{Harrison1978, Harrison1980}. Neugebauer also made significant contributions to these methods \cite{HoenselaersKinnersley1979A, HoenselaersKinnersley1979B, Neugebauer1980A, Neugebauer1980B}.

Sibgatullin's method \cite{Sibgatullin1984} has become widely used for generating exact solutions, with applications by authors such as V.S. Manko \cite{MankoMartinRuiz1994, MankoEtal1994}, L. Herrera \cite{HerreraManko1992}, and E. Ruiz \cite{RuizMankoMartin1995}. Ruiz further refined the method to provide general expressions applicable in a standardised and straightforward manner \cite{MankoEtal1994}. This technique constructs solutions to the Ernst equation based on the form of the Ernst potential along the symmetry axis. Static and axisymmetric vacuum solutions have been derived using relativistic multipole moments, as demonstrated by Hernández-Pastora and Martín \cite{HernandezpastoraMartin1993, HernandezpastoraMartin1994}. These solutions are part of the Weyl family, showcasing the versatility of modern methods in exploring complex spacetime geometries.

In this work, we develop a new method to solve Einstein's equations for stationary, axially symmetric sources by assuming the separability of a metric function, $\Phi(r,\theta)~=~\Phi_R (r) \Phi(\theta)$ and that the spacetime admitting a Killing Tensor field. The separability yields two distinct physically viable solutions: one expressed in polar coordinates and the other involving hyperbolic functions. To ensure consistency, we required a Killing tensor field which generalises the Carter constant\cite{OspinoHernandezpastoraNunez2022} and allows for the separability of additional metric function. By applying the method in polar coordinates, we recover the Kerr solution. This scheme naturally leads to developing a rotating exterior solution for hyperbolic spacetimes.

Hyperbolic geometry in General Relativity enables alternative black hole interior descriptions, preserving staticity and avoiding singularities. It models exotic matter with negative energy density and vacuum cavities, supports thermodynamic insights, and reveals repulsive gravitational effects. Its role extends to dynamic systems, anisotropic compact objects, and non-standard cosmological structures \cite{HerreraDiPriscoOspino2021B, HerreraDiPriscoOspino2021}.

This work is organized as follows. Section \ref{TetradandKinematicalvaribles} introduces the tetrad used to describe the stationary, axially symmetric solution and the scalars derived from the metric. The kinematical variables are then expressed in terms of these scalars. Next, in section \ref{VacuumEquations}, the vacuum field equations regarding the scalars derived from the Ricci identities are explicitly presented. Additionally, specific combinations of these equations are performed to isolate the independent equations that need to be solved. The following section, \ref{GeneralSolution}, focuses on obtaining the general solution to the equations and describes the overall method. It includes resolving static and stationary solutions for polar and hyperbolic geometries. The simplifications introduced by requiring the existence of a Killing tensor are explored in Section \ref{KillingTensor}. Finally, section \ref{FinalRemarks} concludes with a summary of the objectives achieved and key findings.

\section{Tetrad \&  kinematical variables}
\label{TetradandKinematicalvaribles}
We shall consider stationary and axially symmetric sources with the line element written as
\begin{equation}
{\rm d}s^2=-A^2 {\rm d}t^2 +B^2{\rm d}r^2 +C^2{\rm d}\theta^2 +R^2 {\rm d}\phi^2 +2\omega_3 {\rm d}t \, {\rm d}\phi \, ,
\label{Axisymmetric}
\end{equation}
with $A=A( r, \theta)$, $B=B( r, \theta)$, $C=C( r, \theta)$, $R = R( r, \theta)$, and $\omega_3~=~\omega_3( r, \theta)$.

In this case the components of the orthonormal tetrad {\bf V}, {\bf K}, {\bf L} and {\bf S} are:
\begin{eqnarray}\label{TetradGen}
  V^\alpha&=&\left(\frac{1}{A},0,0,0\right), \quad
  K^\alpha = \left(0,\frac{1}{B},0,0\right),\nonumber  \\
   \quad
   L^\alpha & = & \left(0,0,\frac{1}{C},0\right),  \quad {\rm and} \quad S^\alpha=\frac{1}{ \Phi }\left(\frac{\omega_3}{A},0,0,A\right)\, ,\nonumber
\end{eqnarray}
with  $\Phi=\sqrt{A^2 R^2+\omega^2_3}$.

With the above tetrad (\ref{TetradGen}), we shall also define the corresponding directional derivative operators
\begin{equation}
\label{DirectionalDerivatives}
f^{\bullet} = V^{\alpha} \partial_{\alpha}f , \; 
f^{\dag} = K^{\alpha} \partial_{\alpha}f ,  \;
f^{\ast} = L^{\alpha}\partial_{\alpha}f \; \mathrm{and} \;
f^{\circ} = S^{\alpha}\partial_{\alpha}f.
\end{equation}
Additionally, the covariant derivatives of the orthonormal tetrad are
\[
V_{\alpha;\beta}=-a_\alpha V_\beta+\Omega_{\alpha\beta},
\]
\[
 K_{\alpha;\beta}=-a_1 V_\alpha V_\beta+2\Omega_2  V_{(\alpha}S_{\beta)}
+(J_{1}K_\beta+J_{2}L_\beta)L_\alpha+J_{6} S_\alpha S_\beta ,
\]
\[
L_{\alpha;\beta}=-a_2 V_\alpha V_\beta+2\Omega_3 V_{(\alpha}S_{\beta)}
-(J_{1}K_\beta+J_{2}L_\beta)K_\alpha+J_{9} S_\alpha S_\beta ,
\]
and
\[
S_{\alpha;\beta}=-2\Omega_2 V_{(\alpha}K_{\beta)}-2\Omega_3 V_{(\alpha}L_{\beta)}-(J_{6}K_\alpha+J_{9}L_\alpha)S_\beta \, .
\]
The scalars for the axisymmetric metric (\ref{Axisymmetric})  are:
\begin{eqnarray}
  a_1 &=&\frac{A_{,r}}{AB} \, , \, a_2= \frac{A_{,\theta}}{AC} \, , \quad
   J_{1} =-\frac{B_{,\theta}}{BC} \, , \, J_{2}= \frac{C_{,r}}{BC} \, , \\
  J_{6} &=& -a_1+\frac{\Phi_{, \, r}}{B\Phi} \, , \quad
  J_{9} = -a_2+\frac{\Phi_{,\theta}}{C\Phi} \, , \\
  \Omega _2&=& \frac{A^2\psi _{,r}}{2 B\Phi} \quad {\rm and} \quad
\Omega _3= \frac{A^2\psi _{,\theta}}{2 C\Phi}\, ,\label{escOmegas}
\end{eqnarray}
where $\psi = \psi(r,\theta)$ is a general function which allows us to define $\omega_3=A^2 \psi$. 

The kinematical variables $a_\alpha$ and  $\Omega_{\alpha\beta}$ can be written, in terms of the tetrad, as follows
\begin{eqnarray}
a_\alpha&=&a_1K_\alpha +a_2L_\alpha \quad {\rm and}\, , \\
 \Omega_{\alpha \beta}&=&\Omega_2(K_\alpha S_\beta-K_\beta S_\alpha)+\Omega_3 (L_\alpha S_\beta-L_\beta S_\alpha) \, ,
\end{eqnarray}


\section{Vacuum equations}
\label{VacuumEquations}
Now, from  the Ricci identities
\begin{equation}\label{IRicci}
e^{\alpha}_{;\beta;\alpha}-e^{\alpha}_{;\alpha;\beta}=R_{\delta\beta}e^\delta \, , \qquad e\in (\mathbf{V},\mathbf{K},\mathbf{L},\mathbf{S})
\end{equation}
we get
\begin{widetext}
\begin{equation}  \label{va1}
    a_1^\dag+a_2^{\ast} +a_1(a_1+J_{2}+J_{6})+a_2(a_2+J_{9}-J_{1})+2(\Omega_2^2+\Omega_3^2) = 0,
 \end{equation}
\begin{equation}  \label{va2}
\Omega^\dag _2+\Omega_2(2a_1+J_{2})+\Omega^{\ast}_3+\Omega_3(2a_2-J_{1})=0,
 \end{equation}
\begin{equation} \label{va3}
    J_{1}^{\ast}-(a_1+J_{2}+J_{6})^\dag+J_{1}(a_2+J_{9}-J_{1})-a^2_1-J_{2}^2-J_{6}^2+2\Omega_2^2=0,
  \end{equation}
\begin{equation} \label{va4}
    (a_1+J_{6})^{\ast}+a_2(a_1-J_{2})+J_{9}(J_{6}-J_{2})-2\Omega_2\Omega_3=0,
 \end{equation}
\begin{equation}  \label{va5}
    (a_2+J_{9})^\dag+a_1(a_2+J_{1})+J_{1}J_{6}+J_{6}J_{9}-2\Omega_2 \Omega_3=0,
 \end{equation}
  \begin{equation}
   \label{va6}
     J_{2}^\dag +(a_2-J_{1}+J_{9})^{\ast}+a_1 J_{2}+J_{2}J_{6}+a_2^2+J_{1}^2+J_{2}^2+J_{9}^2-2\Omega_3^2=0, \end{equation}
   \begin{equation}  \label{va7}
 J_{6}^\dag+J_{9}^{\ast}+J_{6}(a_1+J_{2}+J_{6})+J_{9}(a_2-J_{1}+J_{9})-2(\Omega_2^2+\Omega_3^2)=0,
\end{equation}
\end{widetext}

The equation (\ref{va1}) can be written as
\begin{equation}\label{va1c}
  (a_1 C\Phi)_{, \, r}+(a_2 B\Phi)_{,\theta}+2BC\Phi (\Omega ^2_2+\Omega_3^2)=0\, ,
\end{equation}
likewise, the equation (\ref{va7}) turns
\begin{equation}\label{va2c}
  (J_{6} C\Phi)_{, \, r}+(J_{9} B\Phi)_{,\theta}-2BC\Phi (\Omega ^2_2+\Omega_3^2)=0 \, ,
\end{equation}
where the corresponding derivatives are denoted by
\begin{equation}
    (\bullet )_{,\, r} \equiv \frac{\partial (\bullet)}{\partial r} \quad {\rm and } \quad
(\bullet )_{,\, \theta} \equiv \frac{\partial (\bullet)}{\partial \theta} \, .
\end{equation}

Next, combining equations (\ref{va1c}) and (\ref{va2c}) we find the following equation for the function $\Phi$
\begin{equation}\label{ecPhi}
  \Phi_{, \, r \, r}+\frac{f_{, \, r}}{f}\Phi_{, \, r}+\frac{1}{f^2}\Phi_{,\theta\theta}=0.
\end{equation}
Here, we have set, without loss of generality, that
\begin{equation}\label{CBf}
  C=Bf(r)\, .
\end{equation}

In the same way, the equation turns (\ref{va4})
\begin{equation}
  \frac{\Phi_{, \, r}}{B\Phi}J_{1}-\frac{\Phi_{,\theta}}{C\Phi}J_{2}+a_2a_1+J_{9} J_{6}=2\Omega_2 \Omega_3 \, . \label{eqC}
\end{equation}
Finally equation (\ref{va2}) can be re-written as follows
\begin{equation}\label{va4c}
  (\Omega_2 A^2 C)_{, \, r}+(\Omega_3 A^2B)_{,\theta}=0 \, . 
\end{equation}


\section{General Solution}
\label{GeneralSolution}
Let's start integrating equation (\ref{ecPhi}) assuming $\Phi~=~\Phi_R (r) \Phi_\Theta (\theta)$ separable, Thus we find three general solutions
\begin{eqnarray}
\Phi_1 = ( c_1 \cos{\mu \theta} + c_2 \sin \mu \theta) \mathcal{F}_1(r)\, , \\
\Phi_2 =  ( \tilde{c}_1 \cosh \mu \theta + \tilde{c}_2 \sinh \mu \theta) \mathcal{F}_2(r) \quad {\rm and} \\ 
 \Phi_3 = (\bar{c}_1 \theta +\bar{c}_2) \mathcal{F}_3(r) \, ,
\end{eqnarray}
where $\mu$ is the separability constant with $c_i$, $\tilde{c}_i$, and $\bar{c}_i$, with $i =1, 2$, are integration constants. If  $\Phi_R (r) = f(r)$, the above three solutions for the equation (\ref{ecPhi}) become particular solutions of the form
\begin{eqnarray}
    \Phi_1 = ( c_1 \cos{\mu \theta} + c_2 \sin \mu \theta) \sqrt{ 2\mu^2 r^2 +c_3 r +c_4}\,  \label{Phimupositiv} \, , \\ 
    \Phi_2 =  ( \tilde{c}_1 \cosh \mu \theta + \tilde{c}_2 \sinh \mu \theta) \sqrt{ -2\mu^2 r^2 +\tilde{c}_3 r +\tilde{c}_4}\,  \, , \label{Phimunegativ} \\ 
    {\rm and} \quad \Phi_3 = (\bar{c}_1 \theta +\bar{c}_2) ( \sqrt{ \bar{c}_3 r +\bar{c}_4} ) \label{Phimuzero}  \, .
\end{eqnarray}
again $c_i$, $\tilde{c}_i$, and $\bar{c}_i$, with $i =1, 2, 3, 4$, are integration constant.   

 The asymptotic behaviour of the metric rules out expressions (\ref{Phimunegativ}) and (\ref{Phimuzero}), but when we do not have this physical restriction, $\Phi_2$ type of solution is also possible. This is the case for the recently discussed examples of hyperbolically symmetric spacetimes~\cite{HerreraWitten2018, HerreraDiPriscoOspino2021}, which will considered below.

Next, we observe that the general solution of (\ref{va4c}) can be written as 
\begin{equation}
  \Omega_2 = \frac{\Omega _{,\theta}}{2A^2C}\qquad {\rm and } \qquad
  \Omega_3 = -\frac{\Omega_{, \, r}}{2A^2B},\label{generalsol}
\end{equation}
where $\Omega = \Omega(r,\theta)$ is an arbitrary function. Now, combining the definition (\ref{escOmegas}) with (\ref{generalsol}) we get
\begin{equation}
\psi_{, \, r} \Omega_{, \, r}+\frac{1}{f^2}\psi_{,\theta} \Omega_{,\theta} = 0 \, . \label{ompsi} 
\end{equation}
Introducing
\begin{equation}\label{eqY}
  Y=\frac{AC}{\sqrt{\Phi}} \, ,
\end{equation}
 equation (\ref{eqC}) becomes
 \begin{equation}\label{eqYd}
   \frac{\Phi_{, \, r}}{\Phi}\frac{Y_{,\theta}}{Y}+\frac{\Phi_{,\theta}}{\Phi}\frac{Y_{, \, r}}{Y}=\frac{2A_{, \, r} A_{,\theta}}{A^2}-\frac{A^4\psi_{, \, r} \psi_{,\theta}}{2\Phi^2} \, .
 \end{equation}

Taking into account  the expression of $\Omega_2$ and $\Omega_3$ from (\ref{generalsol}) and (\ref{escOmegas}), we can write the equation (\ref{va1c}) as
\begin{equation}\label{va1cm}
  (a_1 C\Phi)_{, \, r}+(a_2 B\Phi)_{,\theta}+\frac{1}{2}(\psi_{, \, r}\Omega_{,\theta}-\psi_{,\theta}\Omega_{, \, r})=0,
\end{equation}
or
\begin{equation}\label{eqA}
\left(\frac{A_{,r}}{A} f \Phi \right)_{,r}+\left(\frac{A_{,\theta}}{A} \frac{\Phi}{f}\right)_{, \theta}+\frac{1}{2}(\psi_{,r} \Omega_{, \theta}-\psi_{, \theta} \Omega_{,r}) = 0,
\end{equation}

The integration of (\ref{eqA}) leads us to 
\begin{eqnarray}
\frac{A_{,r}}{A} f \Phi + \frac{1}{2}\psi~\Omega_{,\theta}&=&\tilde{\chi}_{,\theta}\label{xprima} \quad {\rm and} \\
\frac{A_{,\theta}}{A} \frac{\Phi}{f} -\frac{1}{2}\psi~\Omega_{, r}&=&-\tilde{\chi}_{, r} \label{xteta}
%
\end{eqnarray}
where $\tilde{\chi}$ is an arbitrary function. 

Next, the integrability condition of the metric function $A$, leads $\tilde X$ to satisfy 
\begin{equation}
\label{eqxtilde}
    \left( \frac{f \tilde{\chi}_{,r}}{\Phi}\right)_{,r} + \left( \frac{ \tilde{\chi}_{,\theta}}{f\Phi}\right)_{,\theta} = 
    \left( \frac{f \psi \Omega_{,r}}{2 \Phi} \right)_{,r} + \left( \frac{ \psi \Omega_{,\theta}}{2f \Phi} \right)_{,\theta} \, .
\end{equation}
On the other hand, combining the equations (\ref{escOmegas}) and (\ref{generalsol}) we get a relation between the two arbitrary functions 
\begin{eqnarray}
    \psi_{,r}&=&\frac{\Omega_{,\theta} \Phi }{A^4 f}, \label{psip}\\
    \psi_{,\theta}&=&-\frac{\Omega_{,r} f \Phi}{A^4}.\label{psit}
\end{eqnarray}
Finally, taking into account equations (\ref{xprima})-(\ref{xteta}), the integrability condition gives
\begin{equation}
    (\Omega_{,r} f \Phi)_{,r} + \left(\Omega_{,\theta } \frac{\Phi}{f} \right)_{,\theta}=4(\tilde{\chi}_{,\theta}\Omega_{,r} -\tilde{\chi}_{,r} \Omega_{,\theta}).\label{xomega}
\end{equation}


At this point, we can see, from equations (\ref{ompsi})-(\ref{xomega}), that the general solution of this system of equations depends on the choice of an arbitrary function $\Omega$ or $\tilde{\chi}$. 

\subsection{The method}\label{metodo}
The method we shall follow can be stated as
\begin{enumerate}
    \item Integrate equation (\ref{ecPhi}). If $\Phi$ is separable, then it is possible to obtain particular solutions of $\Phi$ in the form of (\ref{Phimupositiv})-\ref{Phimuzero};
    \item provide $\Omega$, and from (\ref{ompsi}) obtain $\psi$;
    \item having $\Omega$ and $\psi$ we get the first metric function $A$ from equations (\ref{psip}) and (\ref{psit});
    \item with the definition $\omega_3 = A^2 \psi$, we obtain the second metric coefficient, $\omega_3$;
    \item calculate the third metric function $R=\sqrt{\frac{\Phi ^2-\omega_{3}^2}{A^2}}$, from the $\Phi$-definition;
    \item from equation (\ref{eqYd}) solve the forth metric coefficient $C$;
    \item finally, equation (\ref{CBf}) leads to the last metric function $B$.
\end{enumerate}
In the next section, we shall illustrate the method with two particular selections of $\Omega$. One leads to a general solution for the spherical static case, and the second recovers the Kerr metric. 

\subsection{ The Static Polar Solutions}
For the static case, we choose
\begin{equation}\label{con1}
     \Omega=0,\quad \Rightarrow \quad \psi=0  
   \end{equation}
and a particular solution of (\ref{Phimupositiv}) with the form $\Phi~=~f\sin \theta~=~\sqrt{r^2-2mr}\sin \theta$. Thus, for $X = \ln A$, equations (\ref{eqYd}) and (\ref{eqA}) become
\begin{equation}\label{eqAS}
  (X_{, \, r} f^2)_{, \, r} \sin \theta+(X_{,\theta} \sin \theta)_{,\theta}=0,
\end{equation}
and
 \begin{equation}\label{eqYS}
   \frac{f_{, \, r}}{f}\frac{Y_{,\theta}}{Y}+\frac{\cos \theta}{\sin \theta}\frac{Y_{, \, r}}{Y}=2X_{, \, r} X_{,\theta} \, ,
 \end{equation}
respectively. 

Additionally, it is easy to check that in the new coordinates
\begin{equation}
\rho=\sqrt{r^2-2mr}\sin \theta \quad {\rm and} \quad z=(r-m)\cos \theta \, ,    
\end{equation}
equation (\ref{eqAS}) transforms into the well-known Weyl equation
 \begin{equation}\label{Weq}
   X_{\rho\rho}+\frac{1}{\rho}X_\rho+X_{zz}=0 \, ,
 \end{equation}
having a general solution written as follows
 \begin{equation}
     X= \ln A_{SP} = \int \frac{\tilde{\chi}_{,\theta}}{f^2\sin\theta}dr+\tilde{A}_{SP}(\theta) \, .
 \end{equation}
Here, $\tilde{A}_{SP}(\theta)$ is an arbitrary function while $\tilde{\chi}$ is any solution of the equation
\begin{equation}
 \tilde{\chi}_{, \, r r}+\frac{1}{f^2}(\tilde{\chi}_{,\theta\theta}-\cot \theta \tilde{\chi}_{,\theta})=0 \, .
\end{equation}
 
 On the other hand, changing variables
  \begin{equation}\label{cv1}
    \eta =\ln f \, ,\quad {\rm and} \quad \xi = \ln \sin \theta \, ,
  \end{equation}
  the equation (\ref{eqYS}) can be written as
  \begin{equation}\label{eqYS1}
    \frac{Y_\eta}{Y}+\frac{Y_\xi}{Y}=2X_\eta X_\xi \, ,
  \end{equation}
having a general solution 
  \begin{equation}\label{eqYS2}
    Y=Exp \left [  \int_{\eta_0}^{\eta}\tilde{f}(\tilde{\chi},\xi-\eta+\tilde{\chi})d\tilde{\chi}  \right ]\tilde{\Delta}(\xi-\eta) \, ,
  \end{equation}
    with $\tilde{f}=2X_\eta X_\xi$, and $\tilde{\Delta}$ an arbitrary integration function.
    
\subsection{The Kerr solution}
For this case, we set
\begin{equation}
\Omega_K=\frac{2m a\cos \theta}{r^2+a^2\cos^2\theta}\, .\label{OmegaK}
\end{equation}
From equation (\ref{xomega}), we find
\begin{equation}
    \tilde{\chi}_{K}=-\frac{m(a^2+r^2)\cos \theta}{r^2+a^2\cos^2 \theta} \, ,
\end{equation}
as a particular solution  and from (\ref{ompsi}), we get 
\begin{equation}
\psi_K=\frac{2mar \sin ^2\theta}{r^2-2mr+a^2\cos ^2\theta},\label{psiK}
\end{equation}
Thus, the first metric coefficient emerges from equations (\ref{psip}) and (\ref{psit}) as
\begin{equation}\label{Akerr}
A^2_K=\sqrt{\frac{\Omega_{K,\theta}\sin \theta}{\psi _{K,r}}}=1-\frac{2mr}{r^2+a^2 \cos^2 \theta} \, .
\end{equation}
Next, the metric functions $\omega_{3K}$ and $R_K$ are calculated as follows
\begin{equation}\label{omegaKerr}
\omega_{3K}=A^2 _K\psi_K=\frac{2mar\sin^2 \theta}{r^2+a^2 \cos^2 \theta}
\end{equation}
and
\begin{equation}\label{RKerr}
R_K=\sqrt{\frac{\Phi ^2_K-\omega_{3K}^2}{A^2_K}}=\sqrt{r^2+a^2+\frac{2a^2mr\sin^2\theta}{r^2+a^2\cos^2\theta}}\sin\theta \, .
\end{equation}
On the other hand, in this case, the right-hand side of equation (\ref{eqYd}) vanishes and its general solution can be written as
\begin{equation}
 \frac{\Phi_{, \, r}}{\Phi}\frac{Y_{,\theta}}{Y}+\frac{\Phi_{,\theta}}{\Phi}\frac{Y_{, \, r}}{Y}=0,\quad \Rightarrow Y=F_1\Big(\frac{\sin \theta}{f}\Big)\label{eqYd1}
\end{equation}
where $F_1$ is an arbitrary function of its argument. 

Under the choice of 
\begin{equation}
    F_1=\sqrt{\frac{f}{\sin{\theta}}-\frac{a^2 \sin\theta}{f}} \, ,
\end{equation}
$C_K$ is obtained from (\ref{eqY}) as
\begin{equation}\label{CKerr}
C_K=\sqrt{r^2+a^2\cos^2\theta} \, ,
\end{equation}
and in turn
\begin{equation}\label{BKerr}
B_K=\frac{C_K}{f}=\sqrt{\frac{r^2+a^2\cos^2 \theta}{r^2-2mr+a^2}} \, .
\end{equation}
Now, as was expected, with equations \ref{Akerr}-\ref{BKerr} we obtain the Kerr metric 
In the case of the Kerr metric, it is  
\begin{widetext}
 \begin{equation}
{\rm ds}^2 =- \left(1-\frac{2mr}{\Lambda}\right){\rm d}t^2-\frac{4mar\sin^2\theta}{\Lambda}{\rm d}t {\rm d}\phi +\frac{\Lambda}{r^2-2mr+a^2}{\rm d}r^2+\Lambda{\rm d}\theta^2
+\sin^2\theta \left(r^2+a^2+\frac{2ma^2r \sin^2\theta}{\Lambda}\right){\rm d}\phi^2 \, ,
      \label{Kerrmetric}
\end{equation}
with $\Lambda = r^2+a^2\cos ^2\theta$
\end{widetext}

\subsection{ The Static hyperbolic solutions}
Motivated by recent works in hyperbolic coordinates in the inner region $r<2m$ \cite{HerreraWitten2018, HerreraDiPriscoOspino2021, HerreraDiPriscoOspino2021B, HerreraEtal2023,HerreraEtal2020}, we shall apply the method to recover the hyperbolic $q-$metric \cite{QuevedoToktarbayYerlan2013}. 

Again, to study hyperbolic static solutions, we set
\begin{equation}\label{con1}
     \Omega_{SH}=0,\quad \Rightarrow \psi_{SH}=0 \, .
   \end{equation}
As a particular solution from equation (\ref{Phimunegativ}) we choose $\Phi= f(r) \sinh \theta = \sqrt{-r^2+2mr-a^2} \sinh \theta$, and clearly $X=\ln(A_{SH})$. Thus, considering $\rho=\sqrt{r^2-2mr}\sinh \theta$, and $z=(r-m)\cosh\theta$ , 
equation (\ref{eqA}) turns again into the hyperbolic Weyl equation
\begin{equation}\label{Weqh}
   X_{\rho\rho}+\frac{1}{\rho}X_\rho+X_{zz}=0 \, .
 \end{equation} 
Now, for the case of the $q-$ metric we have that by setting $a_{2}=0$ in (\ref{va1cm}), we obtain   
\begin{equation}\label{coeAqmetric}
A_{SH}^2(r)=\Big(1-\frac{2m}{r}\Big)^{1+q},
\end{equation}
where we have defined $1+q=-\alpha/2m$.

To find $C_{SH}$, the solution of equation (\ref{eqYd}) is given by 
\begin{equation}
Y=F_2\Big(\frac{\sinh \theta}{f}\Big) \equiv F_1 (-\eta) \label{eqYd1}
\end{equation}
where $F_1$ is an arbitrary function given by
\begin{equation}
    F_2= \frac{1}{\sqrt{\eta \left(\eta ^2 m^2+1\right)^{q(q+2) }}}\, .
\end{equation}
Likewise, 
\begin{equation}\label{coeCqmetric}
C_{SH}^2= r^2 \left(1-\frac{2 m}{r}\right)^{-q}\left(\frac{m^2 \sinh^2(\theta) }{2 m r-r^2}+1\right)^{-q (q+2)},
\end{equation}

\begin{equation}\label{coeBqmetric}
B_{SH}^2=\left(1-\frac{2 m}{r}\right)^{-q-1} \left(\frac{m^2 \sinh ^2(\theta )}{2 m r-r^2}+1\right)^{-q (q+2)}\, ,
\end{equation}
and, 
\begin{equation}\label{coeRqmetric}
R_{SH}^2=r^2\left(1 - \frac{2 m}{r}\right)^{-q}\sinh^2(\theta)\, .
\end{equation}
Considering equations (\ref{coeAqmetric}) through (\ref{coeRqmetric}) finally, the hyperbolic $q$-metric is given by
\begin{widetext}
\begin{equation}\label{qmehiperbolica}
ds^2=-\left(1-\frac{2 m}{r}  \right)^{1 + q}dt^2-\left(1-\frac{2 m}{r}\right)^{-q}\left[\left(1+\frac{m^2 \sinh ^2(\theta )}{2 m r-r^2}\right)^{-q (q+2)} \times \left( \left(1 - \frac{2 m}{r}\right)^{-1}dr^2 +r^2 d\theta^2\right)+r^2\sinh^2\theta d\phi^2 \right],
\end{equation}

\end{widetext}

\subsection{Kerr solution in hyperbolic coordinates}
To implement the Kerr solution in hyperbolic coordinates, we choose
\begin{equation}\label{OmegaKerrHyperB}
  \Omega_{KH}=\frac{2am\cosh\theta}{r^{2}+a^2\cosh ^{2}\theta }.
\end{equation}
From equation \ref{ompsi} we get 
\begin{equation}\label{psihi}
  \psi_{KH}=\frac{2mar\sinh^{2} \theta }{2mr-r^2-a^2\cosh^{2}\theta }.
\end{equation}
Next, combining the equations (\ref{escOmegas}) and (\ref{generalsol}) we obtain that the metric function $A_{KH}^2$ is written as
\begin{equation}\label{AH}
A_{KH}^2=\sqrt{\frac{\Omega_{KH,\theta}\sinh \theta}{\psi^\prime_{KH}}}=\frac{2mr-r^2-a^2\cosh^2\theta}{r^2+a^2\cosh^2\theta},
\end{equation}
and the metric functions $\omega_3$ and $R$ as follows 
\begin{equation}\label{omegaH}
\omega_{3KH}=A_{KH}^2\psi_{KH}=\frac{2amr\sinh^2\theta}{r^2+a^2\cosh^2\theta} ,
\end{equation}
and
\begin{equation}\label{RH}
R_{KH}=\sqrt{\left ( r^2+a^2-\frac{2ma^2r\sinh ^2\theta}{r^2+a^2 \cosh^2 \theta}\right )}\sinh \theta
\end{equation}

On the other hand, as in the case of the Kerr solution, the right side of (\ref{eqYd}) vanishes and its solution can be written as follows 
\begin{equation}\label{YH}
Y_{KH}=\sqrt{\frac{f}{\sinh{\theta}}-\frac{a^2 \sinh\theta}{f}},
\end{equation}
then
\begin{equation} \label{CH}
C_{KH}=\sqrt{r^2+a^2\cosh^2\theta}
\end{equation}
Finally, 
\begin{equation}\label{BH}
B_{KH}=\sqrt{\frac{r^2+a^2\cosh^2\theta}{2mr-r^2-a^2}}
\end{equation}
\begin{widetext}
 \begin{equation}
{\rm ds}^2 =- \left(1-\frac{2mr}{\Lambda_{KH}}\right){\rm d}t^2-\frac{4mar\sinh^2\theta}{\Lambda_{KH}}{\rm d}t {\rm d}\phi +\frac{\Lambda_{KH}}{2mr-r^2-a^2}{\rm d}r^2+\Lambda_{KH}{\rm d}\theta^2
+\sinh^2\theta \left(r^2+a^2-\frac{2ma^2r \sinh^2\theta}{\Lambda_{KH}}\right){\rm d}\phi^2 \, ,
      \label{KerrmetricHI}
\end{equation}
with $\Lambda_{KH} = r^2+a^2\cosh ^2\theta$
\end{widetext}

\section{Axially symmetric Killing tensor}
\label{KillingTensor}
In this section, we shall show that those axially symmetric space-times having a Killing tensor should have a more straightforward separable form for the metric component $C^2=C^2( r, \theta)$. 

For the stationary axially symmetric space-time that admits a Killing tensor \cite{OspinoHernandezpastoraNunez2022} $\xi_{\alpha\beta}$ satisfying the Killing equation 
\begin{equation}
\xi_{\alpha\beta;\mu}+\xi_{\mu\alpha;\beta}+\xi_{\beta\mu;\alpha}=0 \label{KillingEq}
\end{equation}
and written in terms of the tetrad vector is given by
\begin{eqnarray}\label{KillingTensor}
  \xi_{\alpha \beta}&=&\xi_{00}V_\alpha V_\beta+\xi_{11}K_\alpha K_\beta+\xi_{22}L_\alpha L_\beta\nonumber\\
  &+&\xi_{33}S_\alpha S_\beta+\xi_{03}(V_\alpha S_\beta+V_\beta S_\alpha).
\end{eqnarray}
Now, from the integration of the Killing tensor, we have the following
\begin{equation}
  \xi_{11}=F_1(\theta),\quad \xi_{22}=F_2(r)\quad \Rightarrow \quad \xi_{11}-\xi_{22}=C^2\nonumber 
\end{equation}
obtaining 
\begin{equation}\label{conC}
    C^2=F_1(\theta)-F_2(r).
\end{equation}
Also, 

$$\left .\begin{array}{ccc}
   \xi_{33}-\xi_{11} &=&\frac{\Phi ^2}{A^2}g_1(\theta)  \\
  \xi_{33}-\xi_{22} &=&\frac{\Phi ^2}{A^2}g_2(r)  
\end{array}\right \} \,\,  \xi_{22}-\xi_{11} =\frac{\Phi ^2}{A^2}(g_1(\theta)-g_2(r)). $$
from where we have 
\begin{equation}
    A^2C^2=\Phi^2 (g_2(r)-g_1(\theta))\label{conAC}
\end{equation}
On the other hand,
$$\left .\begin{array}{ccc}
   \xi_{03} &=&\Phi ^2\left (g_1(\theta)\psi+C_1(\theta)\right )  \\
  \xi_{03} &=&\Phi ^2\left (g_2(r)\psi+C_2(r)\right )  
\end{array}\right \} \,\,  \psi =\frac{C_2(r)-C_1(\theta)}{g_1(\theta)-g_2(r)}, $$
obtaining
\begin{equation}\label{conpsi}
     \psi =\frac{C_2(r)-C_1(\theta)}{g_1(\theta)-g_2(r)}.
\end{equation}
Likewise, $\xi_{00}$, is given by
$$\begin{array}{ccc}
   \xi_{00}+F_1(\theta) &=&A^2\left [g_1(\theta)(\psi+\frac{C_1(\theta)}{g_1(\theta)})^2+h_1(\theta)\right ] \\
   \xi_{00}+F_2(r) &=&A^2\left [g_2(r)(\psi+\frac{C_2(r)}{g_2(r)})^2+h_2(r)\right ]
\end{array} $$
from we found 
\begin{eqnarray}
     \frac{C^2}{A^2}&=& g_1(\theta)\Big(\psi+\frac{C_1(\theta)}{g_1(\theta)}\Big)^2+h_1(\theta)\nonumber \\
                     &-&g_2(r)\Big(\psi+\frac{C_2(r)}{g_2(r)}\Big)^2-h_2(r)\label{conCA}
\end{eqnarray}
In the following, we will examine the definitions for each integration function for the spherical and hyperbolic cases.


Now, for the spherical case and considering the asymptotic flatness boundary condition, from equations \ref{conAC}-\ref{conCA}, we have that

\begin{equation}\label{AssyCong}
   g_{1}(\theta) = -\frac{1}{\sin^2(\theta)}, \quad C_{1}(\theta)=0,
   \end{equation}
also, without loss of generality we choose  \begin{equation}\label{AssyCong}
    F_{1}(\theta)=h_{1}(\theta)=-b^2\sin(\theta)^2.
    \end{equation}
Under this, the expressions \ref{conC}-\ref{conCA} turns

\begin{equation}\label{conCP}
   C^2 =-b^2\sin(\theta)^2-F_{2}(r),
   \end{equation}
   \begin{equation}\label{conACP}
   A^2C^2 =f^2(g_{2}(r)\sin(\theta)^2+1),
   \end{equation}
   \begin{equation}\label{conpsiP}
   \psi=-\frac{C_{2}(r)\sin^2(\theta)}{g_{2}(r)\sin^2(\theta)+1},
   \end{equation}
  
\begin{eqnarray}
     \frac{C^2}{A^2}&=& -\frac{C^2_{2}(r)\sin^2(\theta)}{(g_{2}(r)\sin^2(\theta)+1)^2}-b^2\sin^2(\theta)\nonumber \\
                     &-&\frac{C^2_{2}(r)}{g_{2}(r)(g_{2}(r)\sin(\theta)^2+1)^2}-h_{2}(r).\label{conCAP}
\end{eqnarray}
Next, combining equations \ref{conCP},\ref{conACP},and \ref{conCAP}, we get
\begin{multline*}
(b^4+f^2b^2g_{2}(r))\sin^4(\theta)\\
+(2b^2F_{2}(r)+b^2f^2+h_{2}(r)f^2g_{2}(r))\sin^2(\theta) \\ 
+F_{2}(r)^2 +f^2\Big( \frac{C_{2}(r)^2}{g_{2}(r)}+h_{2}(r)\Big)=0,
\end{multline*}
where it follows that
\begin{equation}\label{g2}
g_{2}(r)=-\frac{b^2}{f^2},
\end{equation}
\begin{equation}\label{c2}
h_{2}(r)=2F_{2}(r)+f^2,
\end{equation}
\begin{equation}\label{c2}
C_{2}(r)=\pm \frac{b (F_{2}(r)+f^2)}{f^2}.
\end{equation}
Now, as we can see each one of the last expressions depends on the function $F_{2}(r)$, and with this, \ref{conCP}-\ref{conpsiP} turns
\begin{equation}
A^2=\frac{b^2\sin^2(\theta)-f^2}{b^2\sin^2(\theta)+F_{2}(r)},
\end{equation}
and 
\begin{equation}
\psi=\pm\frac{b(F_{2}(r)+f^2)\sin^2(\theta)}{(-b^2\sin^2(\theta)+f^2)},
\end{equation}
and
\begin{equation}
C^2=-b^2\sin^2(\theta)-F_{2}(r).
\end{equation}
Now, for the Kerr metric, the $F_{2}(r)$ is defined by
\begin{equation}
F_{2}(r)=-a^2-r^2,\ \text{where}\ b^2=a^2. 
\end{equation}
\

Likewise, motivated by the approach presented \cite{HerreraWitten2018} and the version of the Schwarzschild solution inside the horizon, 
we must consider this last solution as a limit case for the metric \ref{KerrmetricHI}, when $a=0$, and also well behaved when $r\rightarrow\infty$.

\begin{equation}\label{AssyCong1}
   g_{1H}(\theta) = \frac{1}{\sinh^2(\theta)}, \quad C_{1H}(\theta)=0,
   \end{equation}
   \begin{equation}\label{AssyCong2}
    F_{1H}(\theta)=h_{1H}(\theta)= b^2\sinh(\theta)^2.                \end{equation}
For this case, the expressions \ref{conAC}-\ref{conCA} turns

\begin{equation}\label{conCH}
   C^2 =b^2\sinh(\theta)^2-F_{2H}(r),
   \end{equation}
   \begin{equation}\label{conACH}
   A^2C^2 =f_{H}^2(g_{2H}(r)\sinh(\theta)^2-1),
   \end{equation}
   \begin{equation}\label{conpsiH}
   \psi_{H}=-\frac{C_{2H}(r)\sinh(\theta)^2}{(g_{2H}(r)\sinh(\theta)^2-1)},
   \end{equation}
  
\begin{eqnarray}
     \frac{C^2}{A^2}&=& \frac{C^2_{2H}(r)\sinh(\theta)^2}{(g_{2H}(r)\sinh(\theta)^2-1)^2}+b^2\sinh(\theta)^2\nonumber \\
                     &-&\frac{C^2_{2H}(r)}{g_{2H}(r)(g_{2H}(r)\sinh(\theta)^2-1)^2}-h_{2H}(r).\label{conCAH}
\end{eqnarray}
Next, combining equations \ref{conCH},\ref{conACH},and \ref{conCAH}, we get
\begin{multline*}
(b^4-f_{H}^2b^2g_{2H})\sinh(\theta)^4\\
+(-2b^2F_{2H}+b^2f_{H}^2+h_{2H}f_{H}^2g_{2H})\sinh(\theta)^2 \\ 
+F_{2H}^2 -f_{H}^2\Big( \frac{C_{2H(r)}^2}{g_{2H}}+h_{2H}(r)\Big)=0,
\end{multline*}

where it follows that
\begin{equation}\label{g2H}
g_{2H}(r)=\frac{b^2}{f_{H}^2},
\end{equation}
\begin{equation}\label{c2H}
h_{2H}=2F_{2H}(r)-f_{H}^2,
\end{equation}
\begin{equation}\label{c2H}
C_{2H}(r)=\pm \frac{b (f_{H}^2-F_{2H})}{f_{H}^2}.
\end{equation}
Again, the expression depends on the function $F_{2H}(r)$, and with this, \ref{conCH}-\ref{conpsiH} turns
\begin{equation}
A^2=\frac{b^2\sinh(\theta)^2-f_{H}^2}{b^2\sinh(\theta)^2-F_{2H}(r)},
\end{equation}

\begin{equation}
C^2=b^2\sinh(\theta)^2-F_{2}(r),
\end{equation}
and
\begin{equation}
\psi=\pm\frac{b(f_{H}^2-F_{2H}(r))\sinh(\theta)^2}{(b^2\sinh(\theta)^2-f_{H}^2)}.
\end{equation}
Finally, for the case of the metric in \ref{KerrmetricHI}
\begin{equation}
F_{2H}=-b^2-r^2, \ \text{where}\ b^2=a^2
\end{equation}
On the other hand, for non-rotating configuration, we must recover the metric function related to the spherically static Schwarzschild solution.So, considering $b^2=a^2=0$, we have the following 
\begin{equation}
F_{1}(\theta)=h_{1}(\theta)=C_{2}(r)=g_{2}(r)=0,
\end{equation}
also, 
\begin{equation}
C^2=-F_{2}(r)=r^2,
\end{equation}
and 
\begin{equation}
A^2=\frac{-f^2}{F_{2}(r)}=\Big(1-\frac{2m}{r}\Big).
\end{equation}
The remaining metric functions are obtained by the relationship between $B$ and $C$ and $\Phi$ with $A^2$ and $R^2$.

\section{Final Remarks}
\label{FinalRemarks}
This work presents a method for solving the Einstein equations using the 1+3 tetrad formalism. This approach employs the orthogonal splitting of the Riemann tensor and the covariant derivatives of the tetrad vectors. It transforms the Einstein equations into a set of first-order scalar equations that can be systematically solved, particularly in vacuum, stationary, and axially symmetric systems.

We design a method to solve Einstein Equations for stationary axially symmetric sources.  Assuming that the metric function $\Phi(r,\theta)~=\sqrt{A^2 R^2+\omega^2_3}$ is separable $\Phi(r,\theta)=\Phi_R (r) \Phi$, we integrated equation (\ref{ecPhi}), leading to three distinct solutions: one in polar coordinates, one involving hyperbolic functions, and a linear solution. Next, specifying an arbitrary function $\Omega(r,\theta)$, we integrated the system (\ref{va1})-(\ref{va7}), recovering the well-known Schwarzschild and Kerr solutions. We applied this methodology to the hyperbolic case, reobtaining the static Schwarzschild solution within the horizon in hyperbolic coordinates \cite{HerreraWitten2018}. Extending this approach, we obtained the stationary Kerr solution for analogous physical scenarios \cite{HerreraEtal2020, HerreraDiPriscoOspino2021,HerreraDiPriscoOspino2021B}.

We also discuss the critical role of Killing tensors in identifying and understanding the symmetries of these spacetimes. We show that axially symmetric space-times with a Killing tensor should have a separable metric component $C^2=C^2( r, \theta)$. Killing tensors enable the determination of conserved quantities and simplify the integration of geodesic equations. In rotating and stationary spacetimes, these tensors are particularly valuable, facilitating the separation of variables in the Hamilton-Jacobi and geodesic equations \cite{OspinoHernandezpastoraNunez2022}. This capability results in tractable solutions for particle motion, which are crucial for analyzing phenomena such as gravitational lensing, accretion disk dynamics, and the trajectories of stars and compact objects.

This 1+3 methodology exhibits its robustness and versatility in addressing the complexities of Einstein's field equations for axially symmetric spacetimes. The approach presented here introduces the gauge function, $\Omega=\Omega(r,\theta)$, enabling the derivation of axial stationary exterior solutions for hyperbolic spacetimes. This gauge freedom could also be applied to obtain other Kerr-like solutions where the nonrotating limits would not yield to the standard Schwarzschild scenario. This is ongoing work and will be reported in the future.

\begin{acknowledgments}
L.A.N. acknowledges the financial sponsorship of the Vicerrector\'ia de Investigaci\'on y Extensi\'on de la Universidad Industrial de  Santander and Universidad de Salamanca through the research mobility programs. L.A.N. also thanks the hospitality of the Departamento de Matem\'aticas Aplicadas, Universidad de Salamanca. J.O. and J.L.H.P. express gratitude for financial support from Spain Ministerio de Ciencia, Innovaci\'on, (Programa Estatal de
Generaci\'on de Conocimiento y Fortalecimiento Cient\'\i fico y Tecnol\'ogico del Sistema de I+D+i Grant number: PID2021-122938NB-I00, and Junta de Castilla y Le\'on, (Fondos Feder al $50 \%$ y en l\'\i nea con objetivos RIS3). Grant number: SA097P24. 
\end{acknowledgments}

\bibliographystyle{unsrt}
\bibliography{AllAxially.bib}

\end{document}